\documentclass[preprint]{elsarticle}

\pdfoutput=1

\usepackage{import}

\usepackage{import}
\usepackage{whewell}
\usepackage[hyperindex,breaklinks]{hyperref}
\usepackage{makeidx}
\makeindex

\begin{document}

\begin{frontmatter}
\title{Data Reduction in Deterministic Neutron Transport Calculations Using Machine Learning}

\author{Ben Whewell\corref{au1}}
\ead{bwhewell@nd.edu}
\author{Ryan G.~McClarren\corref{au2}}
\ead{rmcclarr@nd.edu}

\cortext[au1]{Corresponding Author}

\address{Department of Aerospace and Mechanical Engineering \\
University of Notre Dame, Notre Dame, IN 46556, USA
}

\begin{abstract} 
Neutron cross section matrices for fission and scattering data are required for each material, temperature, and enrichment level to calculate the neutron transport equation accurately.
This information can be a limiting factor when using the multigroup discrete ordinates (S$_N$) method when the number of energy groups is large.
Machine Learning (ML) can be used to replace the need for the cross section matrices by reproducing the function that maps the scalar flux to the scattering and fission sources.
Through the use of autoencoders and Deep Jointly-Informed Neural Networks (DJINN), the data storage requirements are reduced by 94\% of the original data for a $618$ group problem.
This is accomplished while preserving the scalar flux, maintaining generality, and decreasing wall clock times.
\end{abstract}


\end{frontmatter}

\section{Introduction} \label{sec:intro}
Reducing the computational costs of neutron transport solvers is a point of interest in the field of reactor physics.
Many of these methods focus on computational speed up ~\cite{Mcclarren:2018,Tano:2019,Sun:2020,Phillips:2021}, but, given the trajectory of high-performance computing architectures with an inexorable reduction in the memory to compute ratio, the topic of data reduction is also of importance \cite{Peng:2019}.
As an example, the recent work of Cherezov et al. takes a data-driven approach to reduce the ``expenses associated with the storing, transfer and processing'' that is part of a multiphysics reactor simulation~\cite{Cherezov:2020}.
In this work, we introduce an approach to embedding the data reduction into the solution of the transport equation by applying machine learning (ML) techniques to predict the action of the fission and scattering matrices in a neutronics calculation.

The multigroup neutron transport equation requires large amounts of storage space for the scattering and fission cross section matrices.
This issue is exacerbated when multiple materials, enrichment levels, temperatures, and scattering moments are used in conjunction with an increased number of energy groups ($G$).
\edit{The nuclear data memory storage can be a limiting factor when applying the neutron transport equation to problems with high numbers of energy groups~\cite{mattingly:2009,jeon:2021, carter:1980, kim:2019}.
}

Our recent work has demonstrated that Deep Jointly-Informed Neural Networks~\cite{humbird2018deep} (DJINN) can be used to reduce the data storage space~\cite{Whewell:2021}, but it can become unfeasible when the number of energy groups is significantly increased.
This problem can be addressed by using a method based on the work of Kluth et al.~\cite{kluth:2020} in high-energy density physics where the input data is first compressed to smaller latent space where the predictions are made before decompressing to the full output space.

In short, we replace the matrix-vector multiplication by a neural network. Each input vector is of size $G$. The construction of this network requires a compression and decompression step that we detail below. Given that we are interested in data reduction, we choose networks that are simple enough to replicate the data without having unnecessary storage requirements.  To train the neural networks, we generate data from a set of nominal transport problems. Our results indicate that our networks are robust to a variety of perturbations in geometry and material content. This will enable analysts to perform parameter searches on a class of problems with a less expensive transport calculation without sacrificing energy fidelity.  

The outline of this paper is as follows; we introduce the concept of replacing the scattering and fission terms in the neutron transport equation in Section 2. 
In Section 3, we demonstrate the use of DJINN on one-dimensional example problems.
This is followed by explaining the implementation of DJINN with autoencoders in Section 4 and the one-dimensional results of a high energy group problem in Section 5.
Section 6 concludes the benefits of using machine learning in the transport equation and discusses additional avenues to pursue.
\section{Transport Calculations with ML Scattering and Fission Reactions} \label{sec:transport}
In this study we consider k-eigenvalue neutron transport problems in one-dimensional geometries under the multigroup approximation~\cite{Lewis:1993}
\begin{equation} \label{eq:transport_equation}
     \mathbf{\widehat{\Omega}} \cdot \nabla \psi_g  (r,\mathbf{\widehat{\Omega}}) + \Sigmatg(r) \psi_g (r,\mathbf{\widehat{\Omega}})
    = \frac{1}{4\pi} \sum_{g'} \Sigma_{\mathrm{s},g' \rightarrow g}(r) \phi_{g'}(r) +
     \frac{1}{4\pi \keff} \sum_{g'} (\nu\Sigma_\mathrm{f})_{g' \rightarrow g}(r) \phi_{g'}(r) \quad g=1,\dots,G.
\end{equation} 
Our notation is standard: $\psig(r,\hOmega)$ is the angular flux for group $g$, $\hOmega \in \mathcal{S}_2$ is angular direction on the unit sphere, $r$ is the spatial variable, and the scalar flux for group $g$ is 
\begin{equation}
    \phig(r) = \int_{4\pi} \psig(r,\hOmega) \, d\hOmega. 
\end{equation}
The material interactions between neutrons and the background medium are characterized by a the total macroscopic cross-section for group $g$, $\Sigmatg(r)$, the scattering macroscopic cross-section from group $g'$ to group $g$ is $\Sigma_{\mathrm{s},g'\rightarrow g}(r)$, and the fission term $(\nu\Sigmaf)_{g'\rightarrow g} \phi_{g'}$ gives the  production rate density of neutrons in group $g$ due to fission occurring in group $g'$.

The scattering and fission terms require the storage of a scattering matrix $\uSigmas$ and fission matrix,  $ \uSigmaf$ matrix, each of size ($G \times G$) where $G$ is the number of energy groups for each material in the problem.
In this work, rather than storing these large matrices for each material variation, we seek to replace the product of a scalar flux vector and the cross section matrices with machine learning models. Using these matrices we can write Eq.~\eqref{eq:transport_equation} as
\begin{equation} \label{eq:transport_equation}
     \mathbf{\widehat{\Omega}} \cdot \nabla \vec{\psi}  (r,\mathbf{\widehat{\Omega}}) + \uSigmat(r) \vec{\psi} (r,\mathbf{\widehat{\Omega}})
    = \frac{1}{4\pi} \uSigmas(r) \vec{\phi}(r) +
     \frac{1}{4\pi \keff} \uSigmaf(r) \vec{\phi}(r),
\end{equation}
where $\vec{\psi}$ and $\vec{\phi}$ are length $G$ vectors containing the angular and scalar fluxes and $\uSigmat$ is the diagonal, total macroscopic cross-section matrix.

In this work, we solve Eq.~\eqref{eq:transport_equation} with the discrete ordinates (S$_N$) method  and the diamond difference spatial discretization~\cite{Lewis:1993}. To solve the resulting equations we used Gauss-Seidel source iteration to converging the scattering source and power iteration to calculate the eigenvalues/vectors. 
For all of the results below, the number of angles ($N = 8$) and the number of spatial cells ($I = 1000$) remained constant.

\subsection{Applying DJINN to Transport Problems}
To use a neural network (NN) in the solution of the neutron transport equation, we will replace the matrix-vector multiplication steps in the source term for both the scattering and fission elements with a NN that takes a scalar flux vector of length $G$ and produces either the scattering or fission source. 
In particular for each power or source iteration, depending on the model, the NN will take the scalar flux at each spatial cell as the input and return the $\uSigmas\vec{\phi}$ or $\uSigmaf(r) \vec{\phi}$ vector for that spatial cell. As an additional input, we can include an indicator value denoting which nuclide mixture is in the cell. We call this a ``labeled'' model: for fuel we use the average atomic weight of the mixture (which in turn indicates the enrichment); in a reflective material we use the molar mass of the scatterer as the label. The labeled models have $G+1$ input parameters.

If our NN models are successful, they will learn to reproduce the matrix vector products needed in the solution of the transport equation and have a smaller storage cost than the necessary data in the original multigroup formulation.

To construct our neural networks, we employ the deep, jointly-informed neural network (DJINN) approach. DJINN was developed to map decision trees to neural networks while also optimizing the hyperparameters of the NNs \cite{humbird2018deep}.
This is a three step process, which constructs a decision tree or random forest from the data, maps this result to a NN or an ensemble of NNs before training the NN.
There are two hyperparameters that the user is concerned about, the number of decision trees and the maximum tree depth.
The number of decision trees change the number of NNs that are created while the maximum depth is correlated to the number of layers in each NN, with the nodes in each layer being related to the number of leaves on the decision tree.
The weights of the NN are initialized using a form similar to Xavier initialization and then tuned through back-propagation. 
The rectified linear unit (ReLU) was used as the activation function and the Adam optimizer was used to minimize the cost function. 
While a number of methods have been used to tune NN hyperparmeters~\cite{snoek,bergstra}, DJINN has been proven to accurately construct a NN without an extensive optimization search in a number of different settings~\cite{Vanderwal:2021,Wang:2020}.
To ensure the DJINN models used were optimized, the number of trees and the maximum depth of the decision trees were varied and tested on the validation setup.

\subsection{Preserving Reaction Rates}
Machine learning is effective at predicting the matrix-vector multiplication but it will not always preserve the reaction rates for both the fission and scattering cross sections.
These reaction rates need to be preserved in order to accurately calculate solutions to eigenvalue problems where any loss of conservation can have a large impact on the estimated eigenvalues and eigenvectors \cite{laboure2017globally}.
To ensure the ML models will preserve overall scattering and fission reaction rates, a scaling factor, \begin{equation} \label{eq:scale}
    m_\ell = \frac{ \sum_{g=1}^{G} \edit{ \sum_{g'=1}^{G} \Sigma_{\ell,g' \rightarrow g} \phi_{g'} } }{\sum_{g=1}^G \left(M_\ell\left(\vec{\phi}\right) \right)_g}, \qquad \ell=\mathrm{s,f},
\end{equation} is used to scale each DJINN predicted value, $M_{\ell}(\vec{\phi})$.
Therefore, the total scattering, $\Sigma_{\mathrm{s},g}$, or fission, $\Sigma_{\mathrm{f},g}$, cross-section for each group is needed; we do not need the scattering or fission matrix. 
\edit{This scaling factor will not ensure that the energy group specific cross sections are correct, but it will preserve the total scattering and total fission reaction rate densities.
Maintaining these reaction rates is necessary when transforming the neutron transport equation from continuous to multigroup energy systems to ensure that the $\keff$ value is preserved~\cite{Demaziere:2019}.
While this results in a spectral shift in lower energy groups, it affects neither the multiplication of the system nor the scalar flux prediction, as we demonstrate below. 
}

\section{$k$-Eigenvalue Results Using DJINN} \label{section:87_group_results}
There were three different materials used in each of the validation problems; a reflective region made up of high density polyethylene (HDPE), a uranium hydride region (UH$_3$) that used a mixture of $^{235}$U and $^{238}$U, and a second uranium hydride region that used only $^{238}$U.
To differentiate between these two uranium hydride regions, they will be referred to as enriched UH$_3$ (with $^{235}$U) and depleted UH$_3$ (without $^{235}$U).
Fudge~\cite{fudge} was used to calculate the multigroup cross sections of each of these materials in which the number of energy groups equals 87.

\begin{figure}[!ht]
    \begin{center}
	\subfigure[]{\includegraphics[]{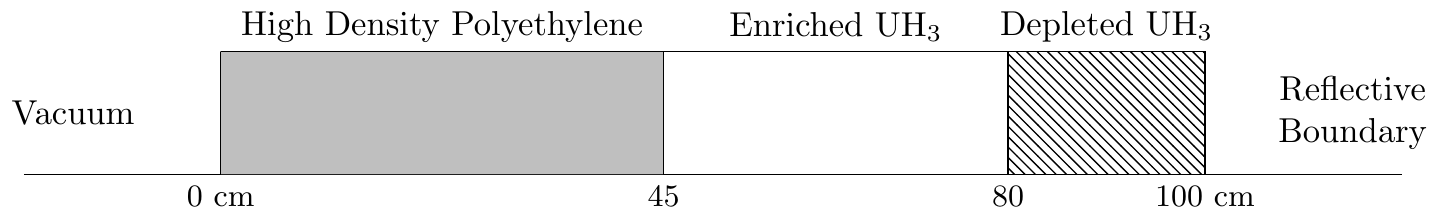}\label{fig:87_group_setup_orig}}
        \subfigure[]{\includegraphics[]{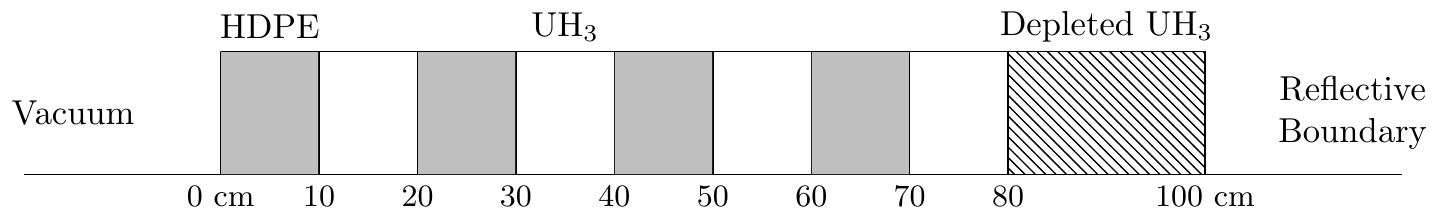}\label{fig:87_group_setup_multi}}
        \caption{Orientation of the one-dimensional problem used for DJINN. In each of these problems, the $^{235}$U enrichment was varied in the enriched UH$_3$ region. (a) This orientation is where the training data was collected. (b) A different orientation is used as a validation test to show generality and how DJINN is able to calculate the interface between the reflective material and the enriched UH$_3$ regions.}
        \label{fig:87_group_setup}
    \end{center}
\end{figure}

There were two different slab orientations used to validate these ML models, as shown in Fig.~\ref{fig:87_group_setup}.
Fig.~\ref{fig:87_group_setup_orig}, which used only three regions, was the original orientation and used for the collection of the training and testing data.
A second orientation of the problem was used, Fig.~\ref{fig:87_group_setup_multi}, with many interfaces between the reflective material and the enriched UH$_3$ regions, in order to show how DJINN is able to apply its findings when there is a major change in the geometry of the problem.
Both of these orientations kept the same one meter width, the same number of spatial cells with a width of one millimeter, and the same boundary conditions with a vacuum boundary on the left and a reflected boundary on the right.

Training data was collected from the material orientation in Fig.~\ref{fig:87_group_setup_orig} at five different enrichment levels: 5\%, 10\%, 15\%, 20\% and 25\% $^{235}$U.
The training data was collected by solving the original problem with discrete ordinates and saving the scalar flux vectors, scattering rate densities, and fission rate densities from the calculations.
In total, there were 275,000 data points for the fission model, 11.44 million data points for the UH$_3$ scattering model, and 9.36 million data points for the HDPE scattering model.
There were no sampling techniques used with the scattering models as the vector length $G$ was small enough to allow for DJINN to train with all data points.
These data points were split among the fission and scattering data as well as between the HDPE and uranium hydride material.
It should be noted that there is no fission data for the HDPE material, so there are three DJINN models: UH$_3$ fission, UH$_3$ scattering, and HDPE scattering.
As detailed in our previous conference paper \cite{Whewell:2021}, it was necessary to build separate models for the scattering in the reflective and fuel materials.

\subsection{Results for the Original Orientation}
\begin{figure}[!ht]
    \centering
    \subfigure[]{\includegraphics[width=0.45\textwidth]{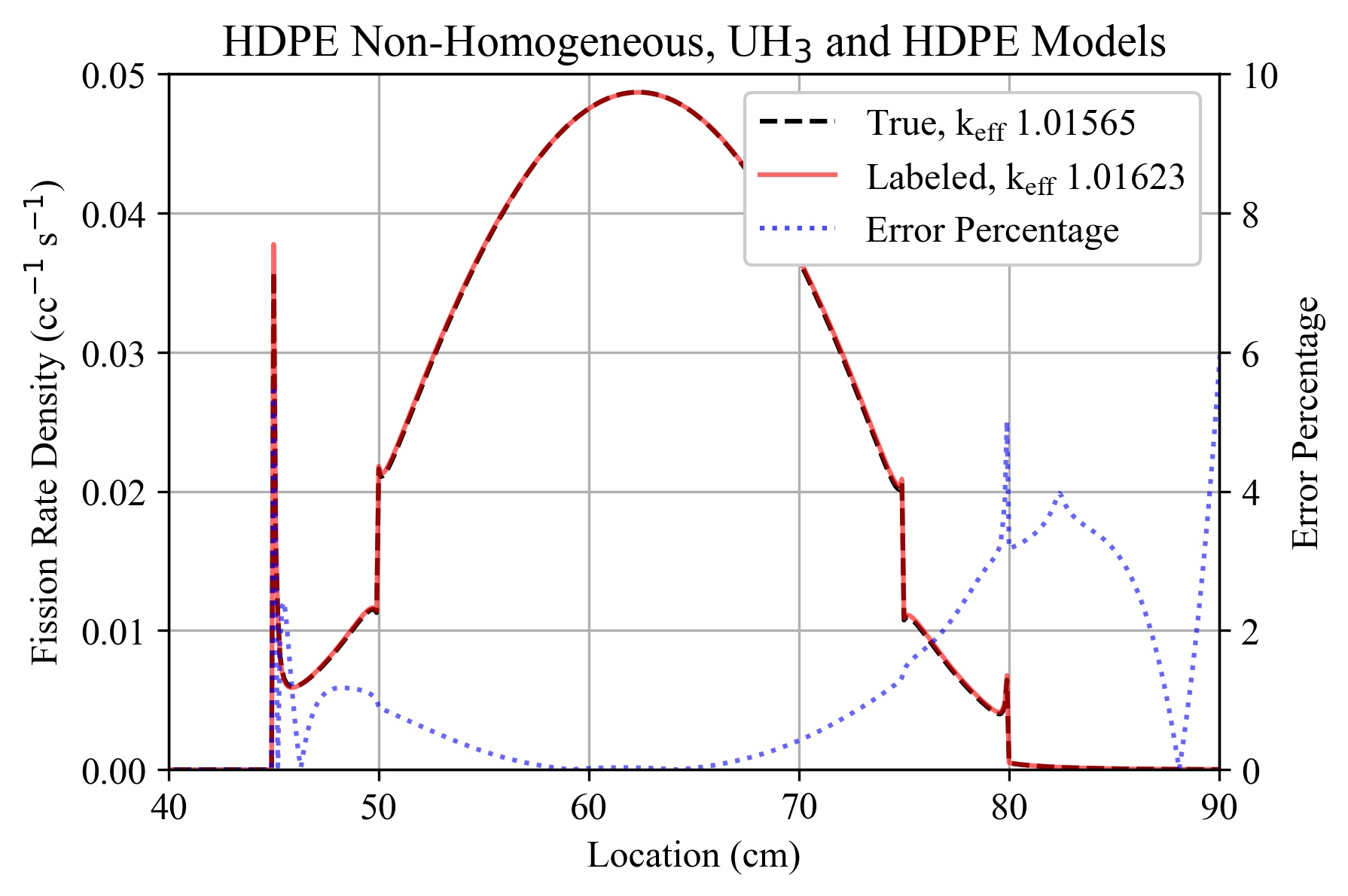}\label{fig:hdpe_mixed_uh3_hdpe_models}} 
    \subfigure[]{\includegraphics[width=0.45\textwidth]{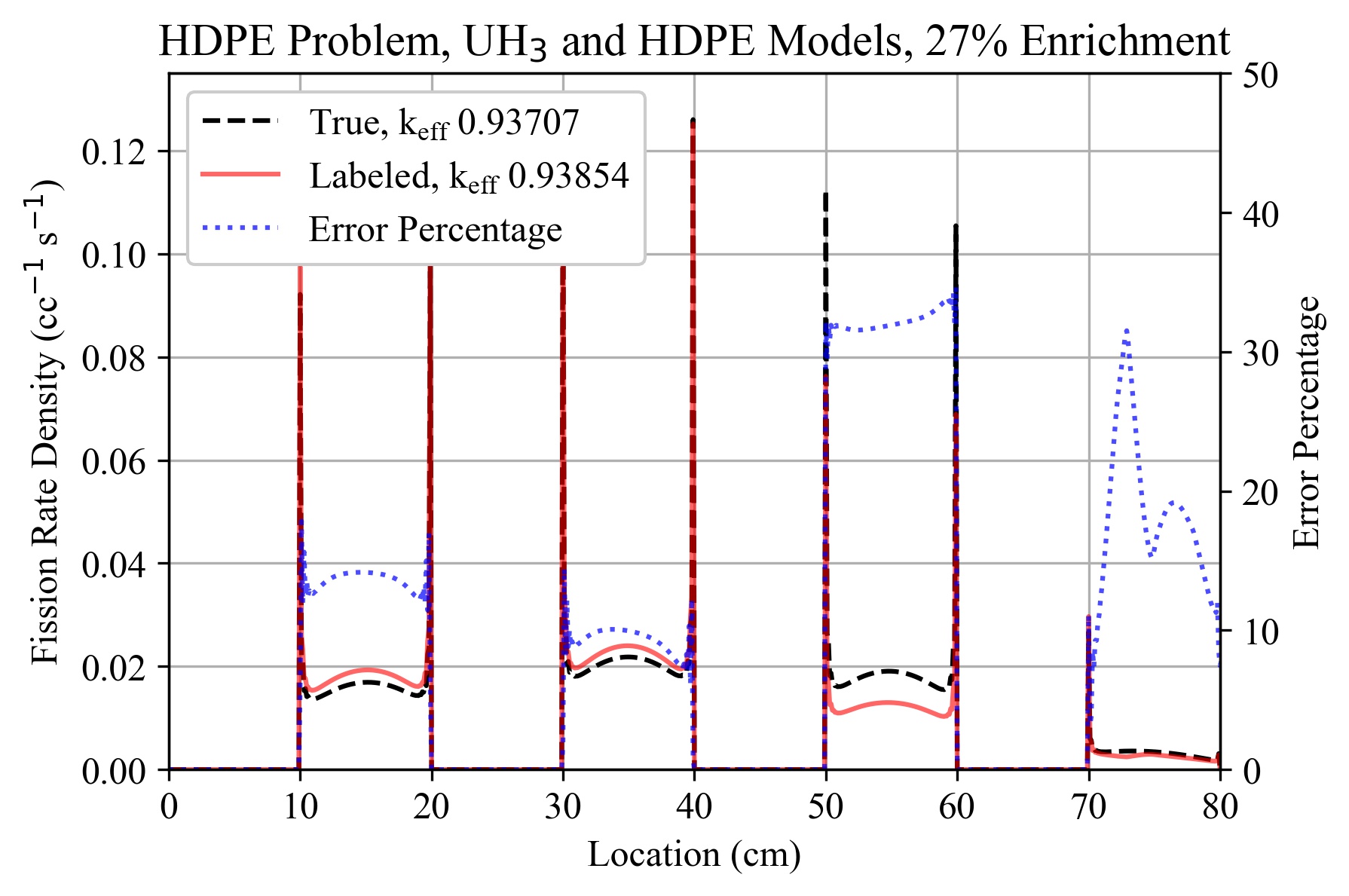}\label{fig:hdpe_multi_uh3_hdpe_models}}
    \caption{Fission Rate Density results for the DJINN UH$_3$ scattering and fission and DJINN HDPE scattering labeled models. \edit{Both these figures include the relative error percentage between the reference solution and the DJINN results.} (a) Shows the result for the original orientation using two different $^{235}$U enrichment levels inside the main fuel region. (b) Shows the result for the multiple interface problem to demonstrate how the model behaves when the geometry is generalized and the enrichment is outside the training data.}
    \label{fig:hdpe_uh3_hdpe_models}
\end{figure}

Validation tests used $^{235}$U enrichments both inside and outside the training data range.
To demonstrate generality with the enrichment variations, a different problem was constructed which used multiple enrichment levels in the enriched UH$_3$ region.
For this case, the first and last five centimeters of the enriched UH$_3$ region used 12\% $^{235}$U and the middle 25 cm used 27\% $^{235}$U.
This differs from the training data, as that used homogeneous enrichment zones composed of a single enrichment level.
Substituting these three DJINN models for the fission and scattering matrices is shown in Fig.~\ref{fig:hdpe_mixed_uh3_hdpe_models}.
This figure shows the fission rate density at each spatial cell; herein we note that DJINN is able to correctly identify the spike at the interface between the reflective material and the enriched uranium hydride. 
We also observe that ML can predict two enrichment levels it had not originally trained on while keeping the $\keff$ within 58 pcm of the reference value.
\edit{This figure also shows the relative error for the fission rate density.
The higher error in the depleted uranium hydride region is related to the lower fission rate density values seen in this region as compared to the enriched uranium hydride region.
There was a higher error at both material interfaces (x = 45, 80 cm), but that was expected and minimal (less than 5\%).}

\subsection{Results for the Multiple Interface Orientation}
To demonstrate that the NN models can be applied to problems with different geometries than used in the training data, we devised a problem with more interfaces between the moderator and the enriched uranium. Moreover, we set the enrichment to 27\%, a value outside the training data.  The result for this extreme case is shown in Fig.~\ref{fig:hdpe_multi_uh3_hdpe_models} in which the DJINN models were able to reproduce the spikes in the fission rate density at the interface of the moderator and fuel.
There was a decrease in the k-effective accuracy, which was within 157 pcm  of the reference value, as compared to the original problem.
\edit{There was also an increase in the relative error, which was significantly higher than observed in Fig.~\ref{fig:hdpe_mixed_uh3_hdpe_models}. 
While the error was the highest at the material boundaries, there was notable error within the uranium hydride regions.}
This was expected as this was an extreme case of the original training data \edit{in which ML was extrapolating for both material and geometry}.
If DJINN models used training data specific to the orientation in Fig.~\ref{fig:87_group_setup_multi}, there would be a decrease in the discrepancy between the reference solution and the DJINN model solution.

\subsection{Data Reduction Results}
The purpose of using ML with the neutron transport equation is to reduce the amount of data required to run multiple problems with slight variations in the material. 
To demonstrate the amount of data storage saved when using the DJINN models instead of standard methods, each fission and cross section matrix for uranium hydride with $^{235}$U enrichments between 0\% and 27\% were stored in a file along with the HDPE scattering matrix.
This was compared to the size of the DJINN models, which included the fission and scatter vectors to preserve the reaction rates.

To choose the correct DJINN models was not obvious, as there is a trade-off between complexity, accuracy, and data storage.
For DJINN, there are two hyperparameters which are chosen, the number of trees and the maximum depth. 
Different combinations of trees and maximum depth were chosen and tested with the validation problem.
An important point to consider is that the fewer trees and maximum depth chosen, the less storage space but also the less complex the model.
While there are certain models that outperformed the models used in Fig.~\ref{fig:hdpe_uh3_hdpe_models}, they would take up 250\% of the original data storage.

\begin{figure}[!ht]
    \centering
    \includegraphics[width=0.7\textwidth]{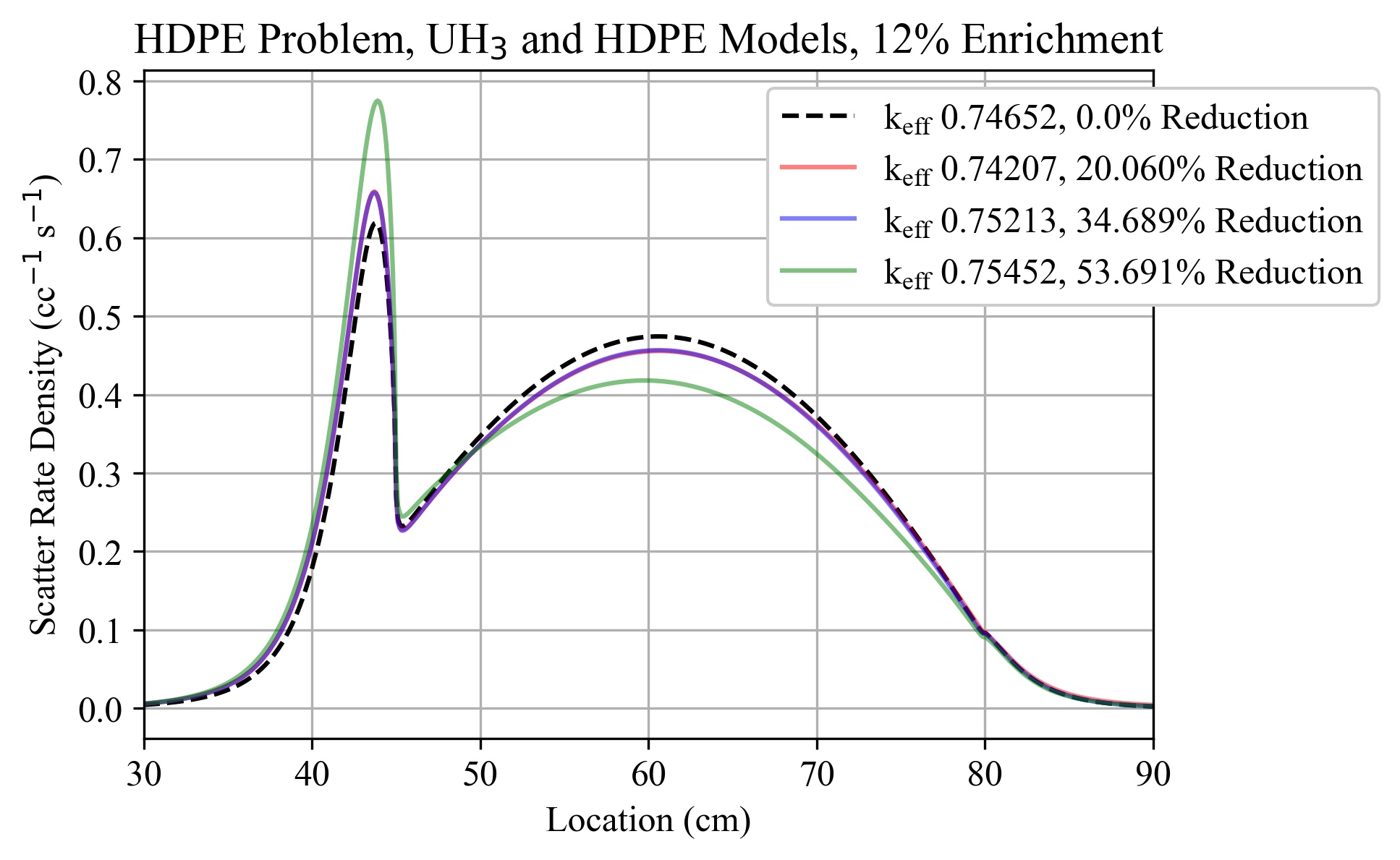}
    \caption{A comparison of the accuracy lost when reducing the amount of data space. This example uses the HDPE non-labeled scattering DJINN model and both of the UH$_3$ non-labeled DJINN models for a 12\% enriched UH$_3$ region. This plot displays the Scattering Rate Density.}
    \label{fig:error_vs_data}
\end{figure}

Instead, the complexity was reduced at the cost of being accurate.
This trade-off between the model complexity and the accuracy of the validation results is shown in Fig.~\ref{fig:error_vs_data}. 
This figure uses the scattering rate density to exemplify the differences in the model complexity and the accuracy.
The simpler model, with a 53.691\% reduction of the original data, is able to reduce additional storage space but it is not as accurate as the other models which reduce the data by 34.689\% and 20.060\%. 

Combining the data space required for the UH$_3$ fission, UH$_3$ scattering, and HDPE scattering ML models in Fig.~\ref{fig:hdpe_uh3_hdpe_models}, there was an 40.682\% reduction.
Given the HDPE scattering model is only predicting a single matrix, this full matrix and ML model can be removed from the analysis to show a data reduction of 59.826\%.
There are more opportunities for data reduction and decreasing the overall matrix storage requirements, such as enrichment values, the inclusion of temperature, and anisotropic scattering matrices. 
\section{Autoencoders for problems with a large number of energy groups}
The application of DJINN is limited by the size of the input vector, namely the number of energy groups used for the multigroup transport code. 
Up until this point, the number of energy groups ($G = 87$) has been relatively small. 
There are problems where the number of energy groups can be in the hundreds.
In these cases, both training and implementing the DJINN models would be difficult and might not be beneficial to the data saving measures.

\begin{figure}[!ht]
    \begin{center}
        \includegraphics[]{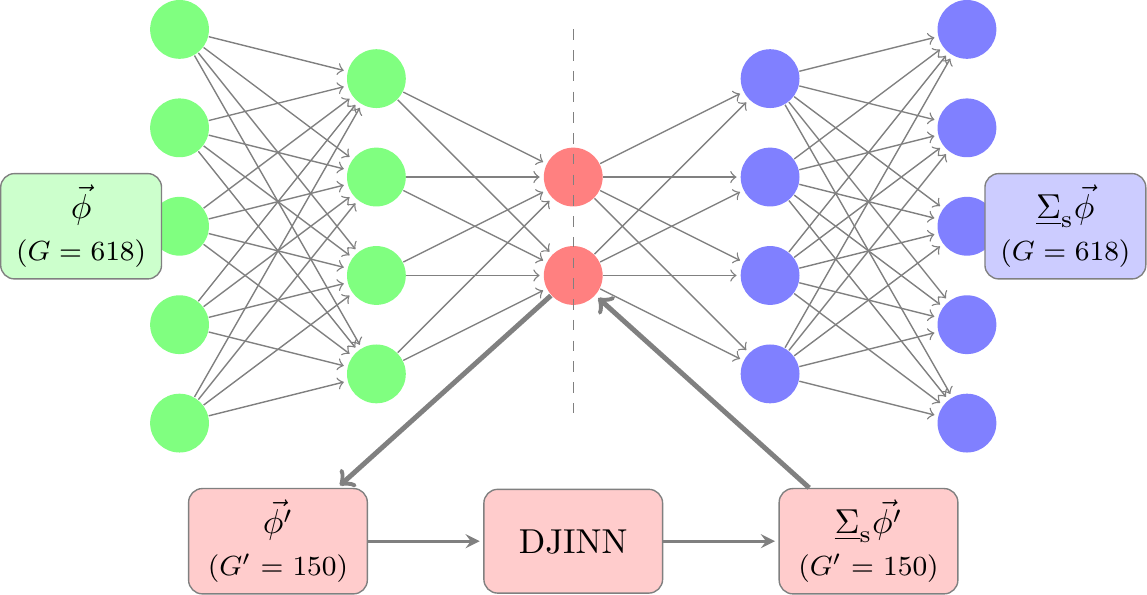}
        \caption{Incorporating DJINN with autoencoders: The scalar flux encoder is shown by the green nodes and the matrix-vector multiplication decoder is shown with the blue nodes. The latent space is shown in red. \edit{The models used three hidden layers (300, 150, 300), with a latent space of $G' = 150$.}}
        \label{fig:autoencoder}
    \end{center}
\end{figure}

One solution is to use autoencoders, a type of neural network, to reduce the input vector of DJINN, thus reducing the issues with training and dealing with the memory overhead.
An autoencoder is comprised of two neural networks, an encoder and a decoder~\cite{Goodfellow:2016}. 
The input vector will pass through the encoder and reduce the size of the vector into a latent space.
The decoder will then attempt to represent the input vector from the data in the latent space.

In order to apply autoencoders with the current DJINN application, two autoencoders will be trained for each DJINN model, one for the scalar flux and a second for the matrix-vector product.
The full autoencoders will be used for training, but, for the implementation, only the encoder of the scalar flux autoencoder and the decoder of the matrix-vector multiplication autoencoder are used.
As shown in Fig.~\ref{fig:autoencoder}, the input scalar flux vector will be passed through the scalar flux encoder and the vector will be reduced from size \edit{$G = 618$ to size $G' = 150$} in the latent space \edit{for this specific case}.
The latent space is where the DJINN model will be trained. 
It will map the latent space of the scalar flux to the latent space of the matrix-vector multiplication.
Finally, the latent space matrix-vector multiplication of size $G'$ will pass through the decoder for the matrix-vector multiplication and return to size $G$.

\subsection{Training the Autoencoders}
\begin{figure}[!ht]
    \begin{center}
	\includegraphics[]{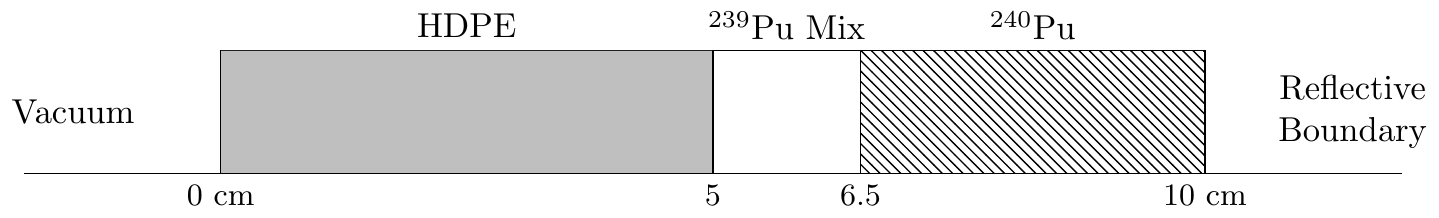}
        \caption{Orientation of the one-dimensional 618 group problem to use for the DJINN-Autoencoder models. The plutonium mixture region was comprised of different percentages of $^{239}$Pu and $^{240}$Pu.}
        \label{fig:618_group_setup}        
    \end{center}
\end{figure}

The one-dimensional problem setup will use HDPE in addition to plutonium-239 and plutonium-240, all with 618 energy groups. 
This is similar to the $G = 87$ group problem however, the ``enrichment'' section will be made up of a mixture of $^{239}$Pu and $^{240}$Pu and the width of the problem will be 10 centimeters. 
The problem is shown in Fig.~\ref{fig:618_group_setup}, with the training data using different $^{239}$Pu percentages (between 75\% and 95\% $^{239}$Pu) in the mixture region.

As before, the data was separated between the reflective material (HDPE) and the fuel material (plutonium) for two different scattering models in addition to the one fission model (plutonium). 
The autoencoders used three fully connected hidden layers of sizes 300, 150, and 300 nodes\edit{, where $G' = 150$}. 
Additional layers were not included to reduce the number of weights required for each model.

The scattering models from both the HDPE and plutonium materials collected 3.7 million data points each from the five different mixtures. 
Instead of attempting to train all of this data for both the scalar flux and matrix-vector multiplication, 150,000 data points were randomly sampled from each of these mixtures, resulting in 750,000 data points with an 80/20 train-test split. 
The number of training iterations (epochs) were varied between 100 and 600 with resampling occurring after every 100 epochs to ensure adequate sampling of the data.
The data was transformed using a cube-root function to compress the dynamic range of the fluxes and matrix-vector products. 
\edit{We also investigated logarithmic and min-max transformations, but the cube-root transformation yielded the most accurate results for both scattering models.}
The performance of each of these autoencoders were based off of results in which the autoencoder was used for both the scalar flux and the matrix-vector multiplication at each location the DJINN-autoencoder operation would be called.
Mixtures of 73\%, 85\% and 88\% $^{239}$Pu were used for these tests.

The autoencoders for the fission models used 36,500 training and testing points with a 80/20 train/test split.
The number of epochs were also varied without the resampling process. 
A difference between the scattering and fission data was that the min-max transformation was used, which required saving the maximum and minimum vectors for the input to be used to reverse transform the output data.
\edit{Both the cube-root and log transformations were used but did not result in models as accurate as the min-max transformation for the plutonium fission model.}
The same method to compare the performance of the autoencoders of the scattering models was used for the fission models.

\subsection{Training the DJINN Models}
For the training of the DJINN models, the original data had to be transformed down to the latent space, so instead of $G = 618$, $G' = 150$ was used as the size of the DJINN input.
The scattering data was sampled evenly between the five mixture levels so 750,000 data points were used for the training and testing.
The same 36,500 fission data points were used for the fission model's training and testing.
All three DJINN models (HDPE scattering, plutonium scattering, and plutonium fission), the number of trees and the maximum depth were varied.
The models were chosen based off their mean squared error, the validation results for the three different mixture levels (73\%, 85\%, and 88\% $^{239}$Pu), and the amount of data saved.
The validation mixture levels were chosen to show the generality of the DJINN-autoencoder application.

\section{$k$-Eigenvalue Results Using Autoencoders and DJINN}
\begin{figure}[!ht]
    \centering
    \subfigure[]{\includegraphics[width=0.495\textwidth]{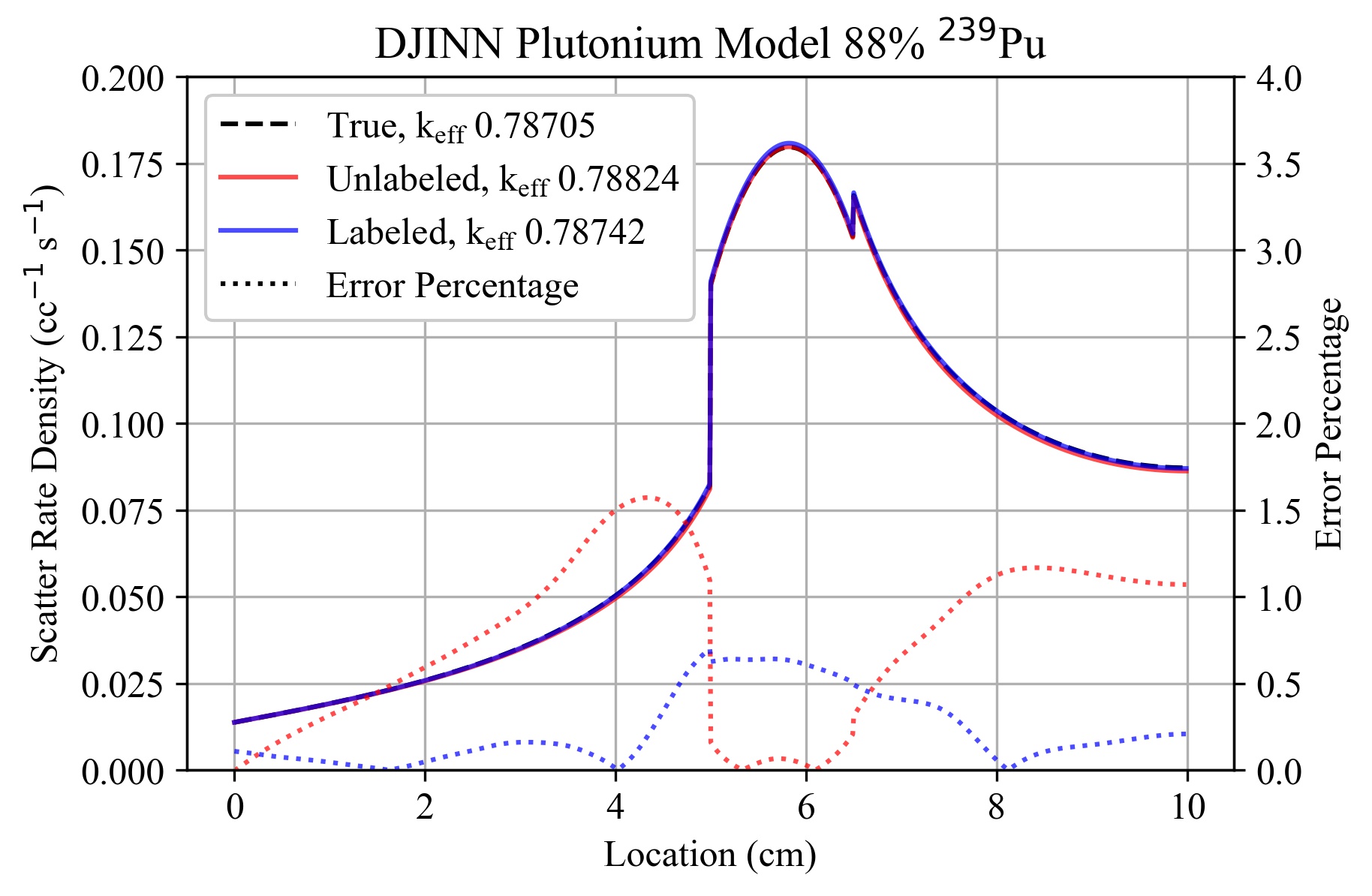}\label{fig:pluto_validation_results_12}}
    \subfigure[]{\includegraphics[width=0.495\textwidth]{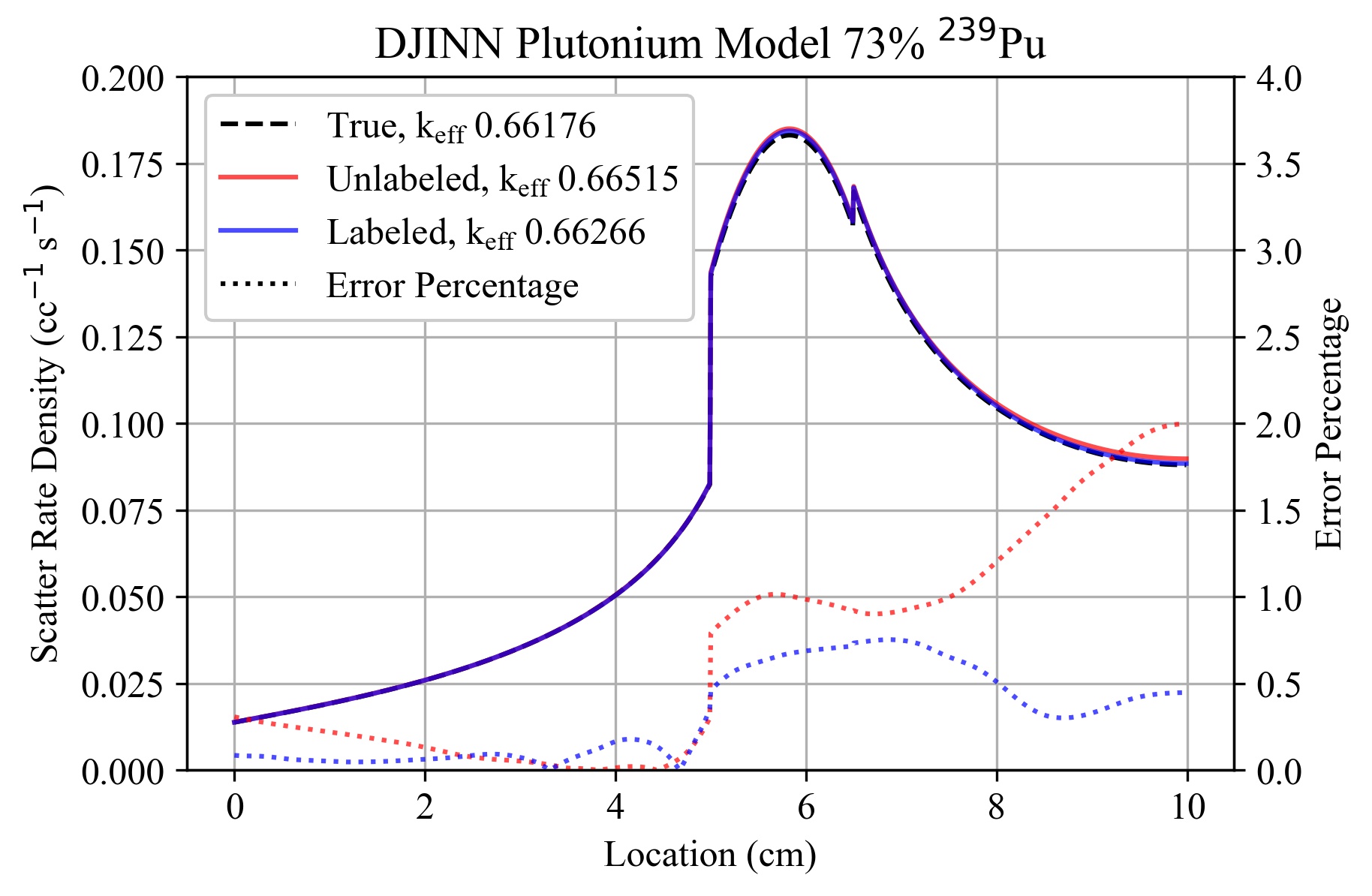}\label{fig:pluto_validation_results_27}}
    \caption{Scattering Rate Density results for the unperturbed orientation using the DJINN-autoencoder plutonium scattering and fission models and the DJINN-autoencoder HDPE model. \edit{The relative error percentage is shown for each model.} (a) Shows a mixture of $^{239}$Pu within the range of the training data. (b) Shows a mixture of $^{239}$Pu outside the training data.}
    \label{fig:pluto_validation_results}
\end{figure}

The results for the 88\% and 73\% $^{239}$Pu mixtures can be seen in Fig.~\ref{fig:pluto_validation_results}. 
These results show that the DJINN-autoencoder models are able to correctly predict the matrix-vector multiplications for models both inside and outside the $^{239}$Pu mixture range of the training data.
Each plot shows the scattering rate density for both the DJINN-autoencoder models with and without a label.
While both the non-labeled and labeled models were accurate, the k-effective value of the labeled model is closer to the reference value.
\edit{This can be observed in the relative error of the figures, where the error in the labeled model is less than the error for the unlabeled model in most instances. 
Overall, the DJINN-autoencoder model errors had less than 2\% error in each region.

These errors are also less than those found in Fig.~\ref{fig:hdpe_uh3_hdpe_models} since there is not a significant change to the material orientation. 
Fig.~\ref{fig:hdpe_mixed_uh3_hdpe_models} has two different enrichment regions and Fig.~\ref{fig:hdpe_multi_uh3_hdpe_models} changes the orientation. 
However, Fig.~\ref{fig:pluto_validation_results} only changes the percentage of the plutonium, which is closer to the training data than the previous results.}

\begin{figure}[!ht]
    \centering
    \subfigure[]{\includegraphics[width=0.495\textwidth]{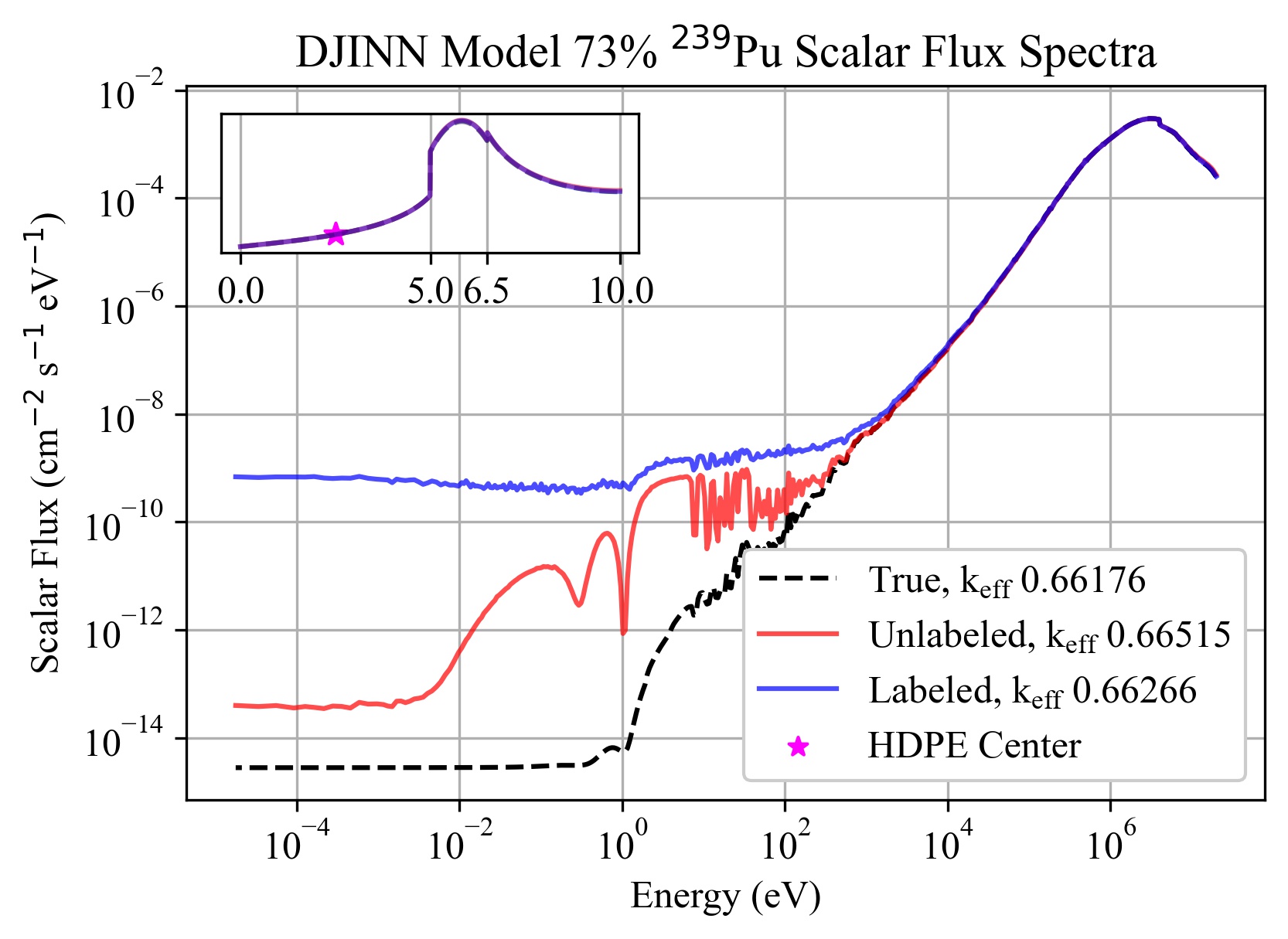}\label{fig:spectra-hdpe-flux}}
    \subfigure[]{\includegraphics[width=0.495\textwidth]{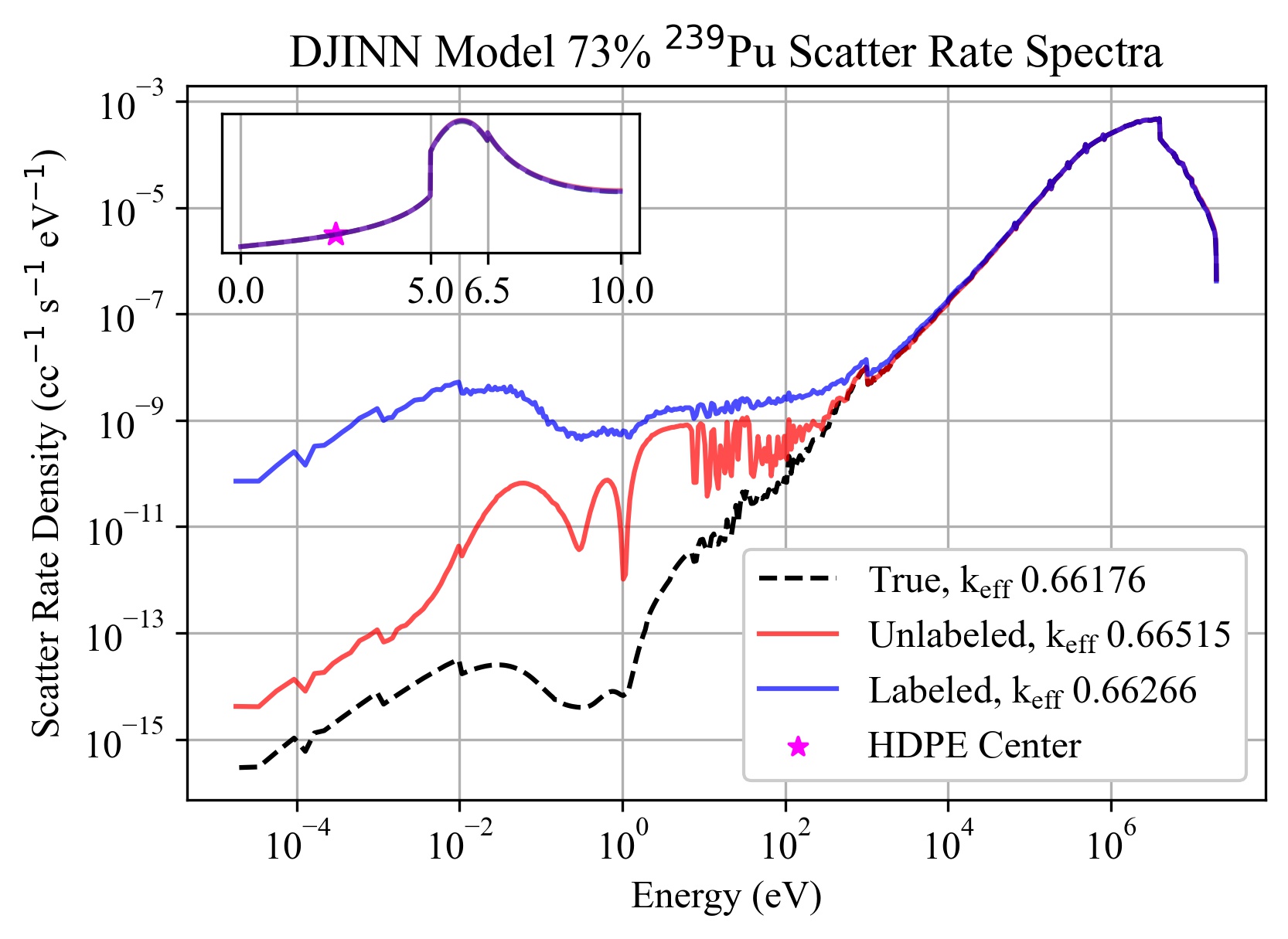}\label{fig:spectra-hdpe-scatter}}
    \subfigure[]{\includegraphics[width=0.495\textwidth]{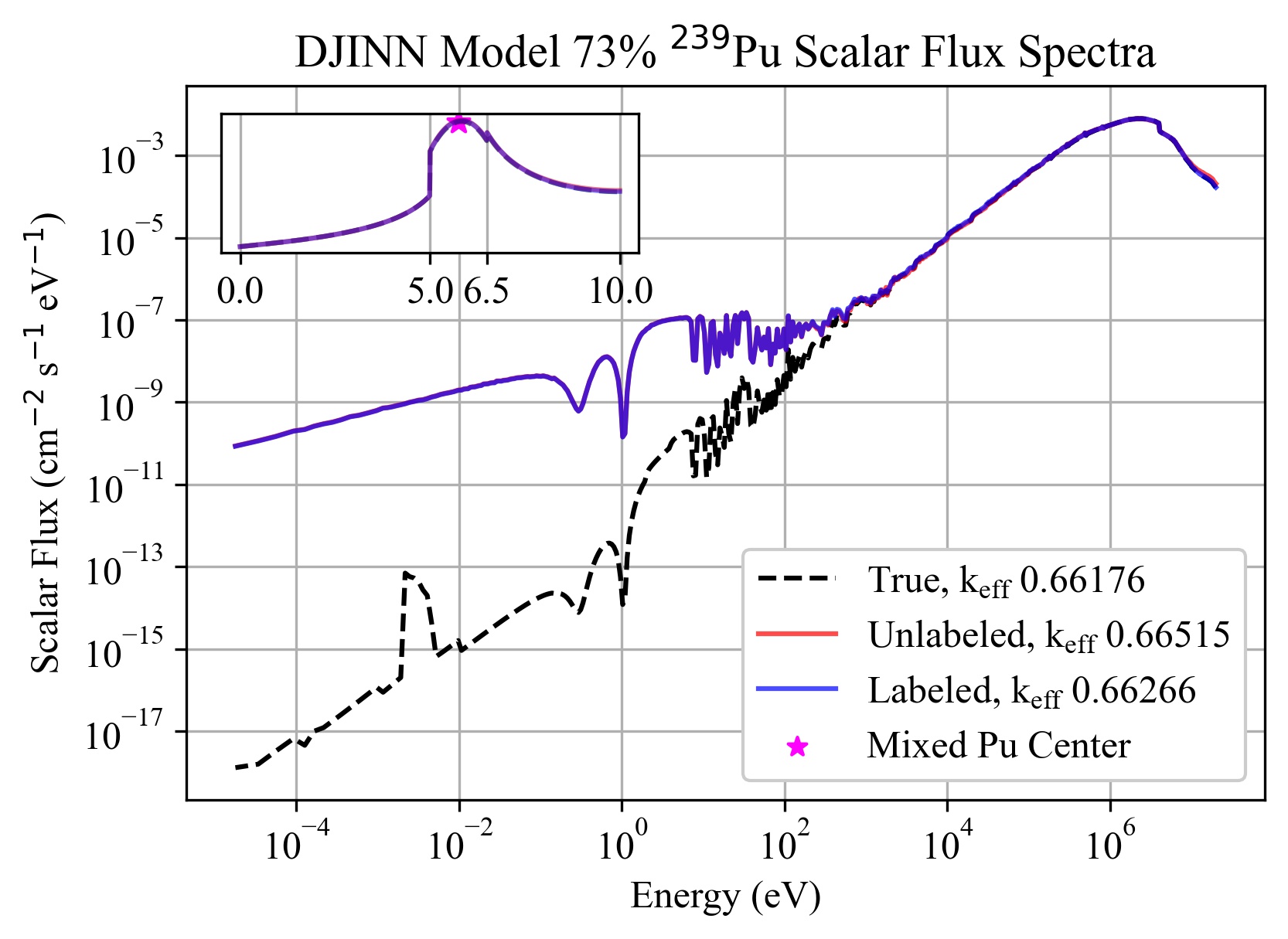}\label{fig:spectra-pluto-flux}}
    \subfigure[]{\includegraphics[width=0.495\textwidth]{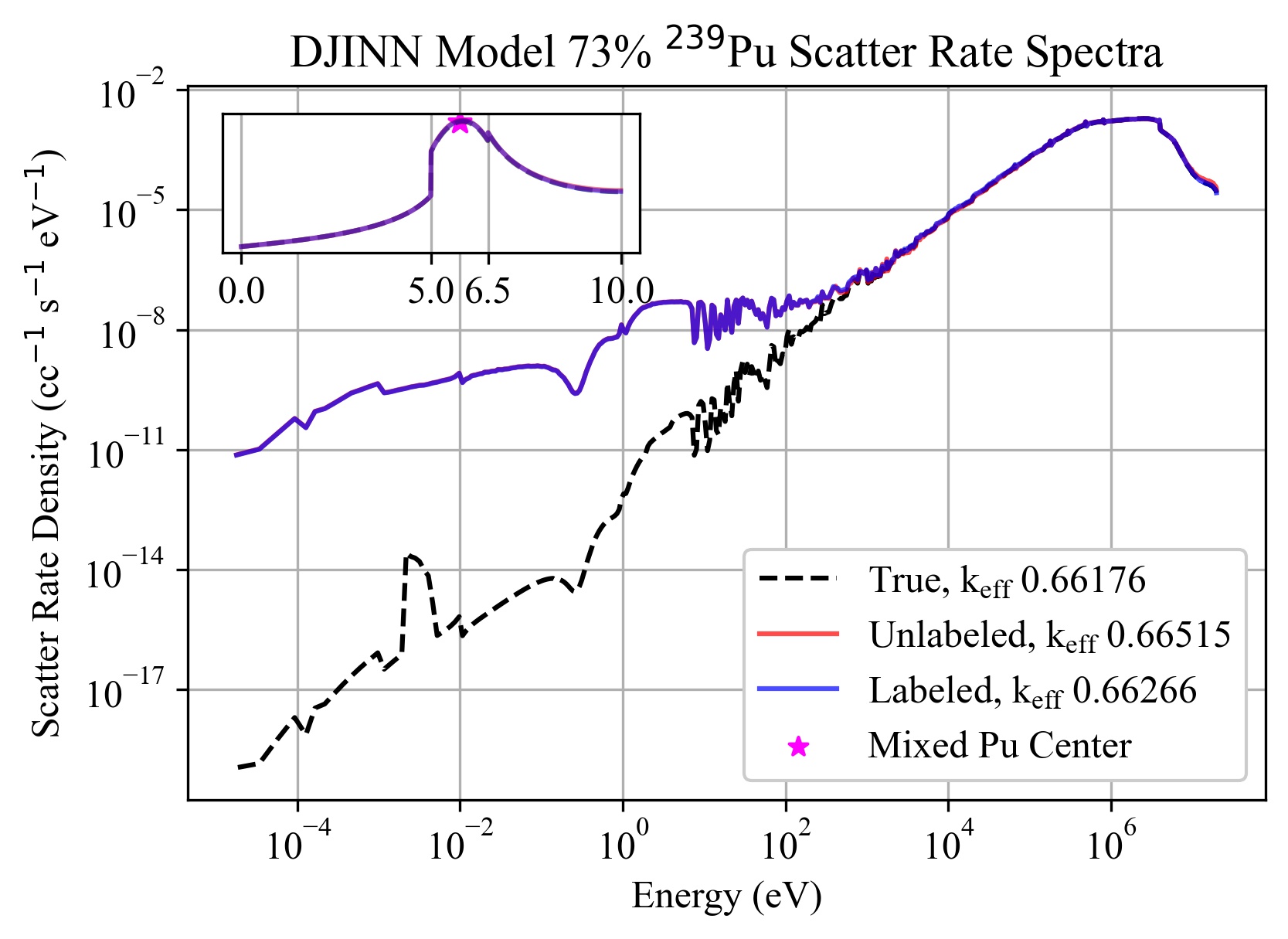}\label{fig:spectra-pluto-scatter}}
    \subfigure[]{\includegraphics[width=0.495\textwidth]{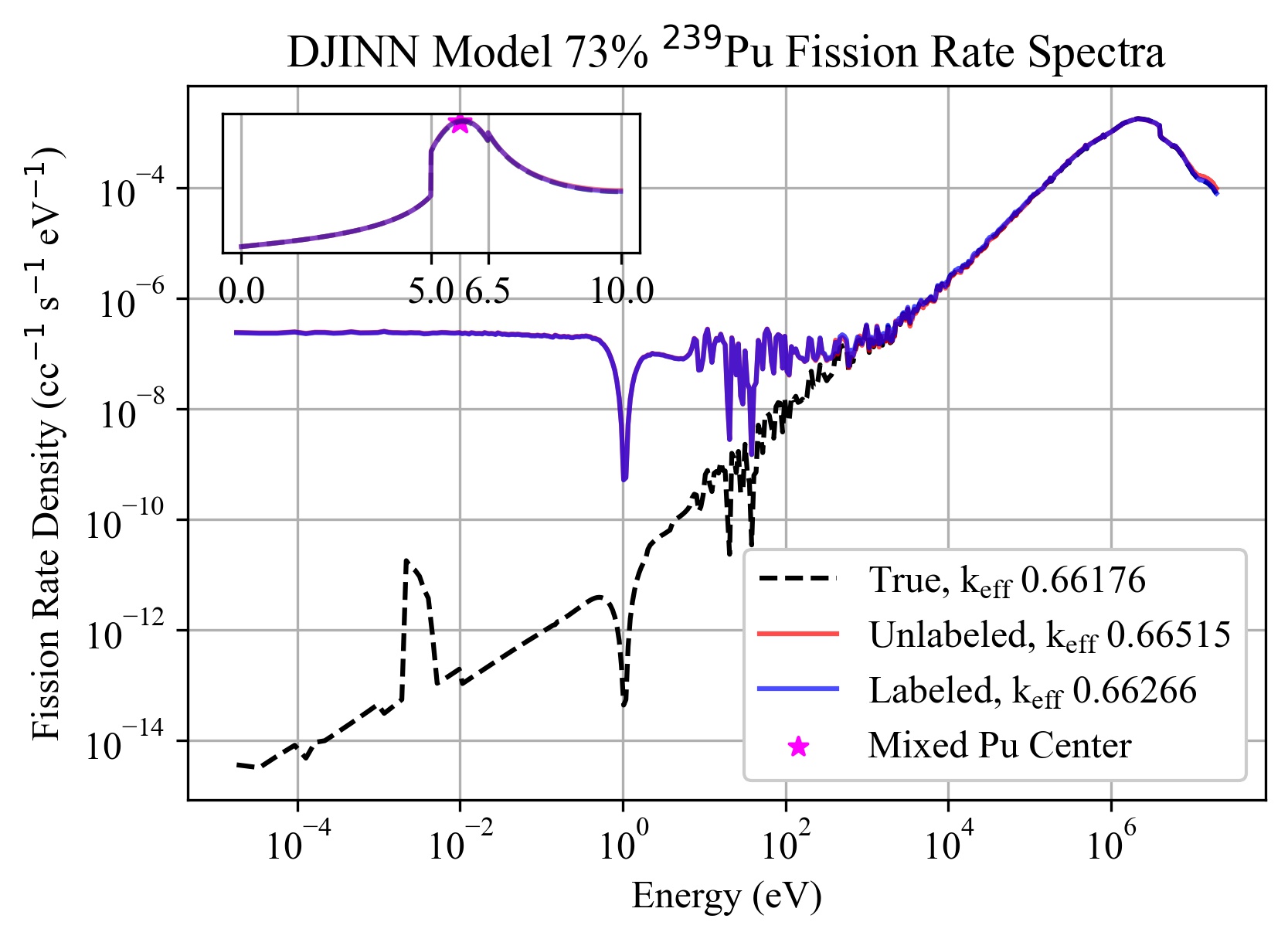}\label{fig:spectra-pluto-fission}}
    \caption{\edit{Different spectra of the plutonium system using autoencoders and DJINN for the problem shown in  Fig.~\ref{fig:pluto_validation_results_27}. While both the unlabeled and labeled models had difficulty in the lower energy regions for all instances, this did not affect the global scalar flux or the $\keff$ results. (a - b) Show the flux and scattering rate spectra for the midpoint of the reflective HDPE region. (c - e) Show the flux, scattering rate, and fission rate spectra for the midpoint of the mixed plutonium region.}}
    \label{fig:pluto-spectra}
\end{figure}

\edit{As mentioned in Section~\ref{sec:transport}, the scaling factor preserves the reaction rates but does not ensure that the group-specific reaction rates are preserved.
While the results in Fig.~\ref{fig:pluto_validation_results} demonstrate that the DJINN-autoencoder models are able to accurately predict the scattering rate density, Fig.~\ref{fig:pluto-spectra} shows a spectral shift for the lower energies. 
This is likely due to the large range of the spectra and was consistent across models that used different data normalization and transforms. We can also confirm that switching the models from single to double precision does not materially affect the results.

Figure~\ref{fig:pluto-spectra} shows the scalar flux, scatter rate density, and fission rate density as a function of energy at specific points for the results from Fig.~\ref{fig:pluto_validation_results_27}.
Figs.~\ref{fig:spectra-hdpe-flux} and~\ref{fig:spectra-hdpe-scatter} show the flux and scatter rate spectra for the midpoint of the HDPE reflective region. 
The DJINN-autoencoder model spectra agree with the reference solution when the energy is above 1 keV but have difficulties in predicting the spectra at lower energy levels. 
The same can be said for Figs.~\ref{fig:spectra-pluto-flux},~\ref{fig:spectra-pluto-scatter}, and~\ref{fig:spectra-pluto-fission}.
This flux, scattering rate, and fission rate spectra also have issues correctly predicting the spectra at the midpoint of the mixed plutonium region below 1 keV.
When the energy is below 1 keV, the spectra is many orders of magnitude lower than values above 1 keV, which becomes difficult for our ML models to capture.
While the magnitude of the DJINN-autoencoder models is inaccurate at lower energies, some of the shapes agree with the reference spectrum, such as the dip in Fig.~\ref{fig:spectra-pluto-fission} at 1 eV. 
The lower spectral shift does not prohibit the accuracy when solving for the $\keff$ values with the ML models. 
For less accurate models, such as Fig.~\ref{fig:pluto_perturbation_results_27}, there is no agreement between the reference spectra and the ML models.
This points to the fact that, to preserve the $\keff$ value, part of the spectrum must be accurate.
}
\newpage

\subsection{Data Reduction Results}
The purpose of using machine learning is to completely replace the storing of the fission and scattering matrices when running the S$_N$ transport code.
The reference data storage is comprised of the 28 fission and scattering matrices for each plutonium mixture with $^{239}$Pu between 0\% to 27\%.
Included in this data storage are the fission and scattering matrix from the $^{240}$Pu material and the scattering matrix from the HDPE.

In this problem, the original nuclear data consisted of a fission spectrum, $\chi$, and a vector of $\nu\Sigmaf$ values for the groups. Therefore, when we compare the storage used in the autoencoders and the DJINN models with the storage of the few vectors needed for the fission data and the nonzeros in the scattering matrices, we still have a notable reduction in the storage size of 25.015\%. 
If, however, we use the matrix forms of the fission data as a reference, the DJINN-autoencoder models and scaling vectors use 6.410\% of the original data.
\edit{We note that the angular and spatial discretization will not affect the memory requirements, except insofar as higher-order scattering components are added. In a sense, our results are conservative because we are only dealing with isotropic scattering: high-order scattering moments will increase the benefit (and necessity) of compression by increasing the memory required for nuclear data.
}

\begin{figure}[!ht]
    \centering
    \includegraphics[width=0.7\textwidth]{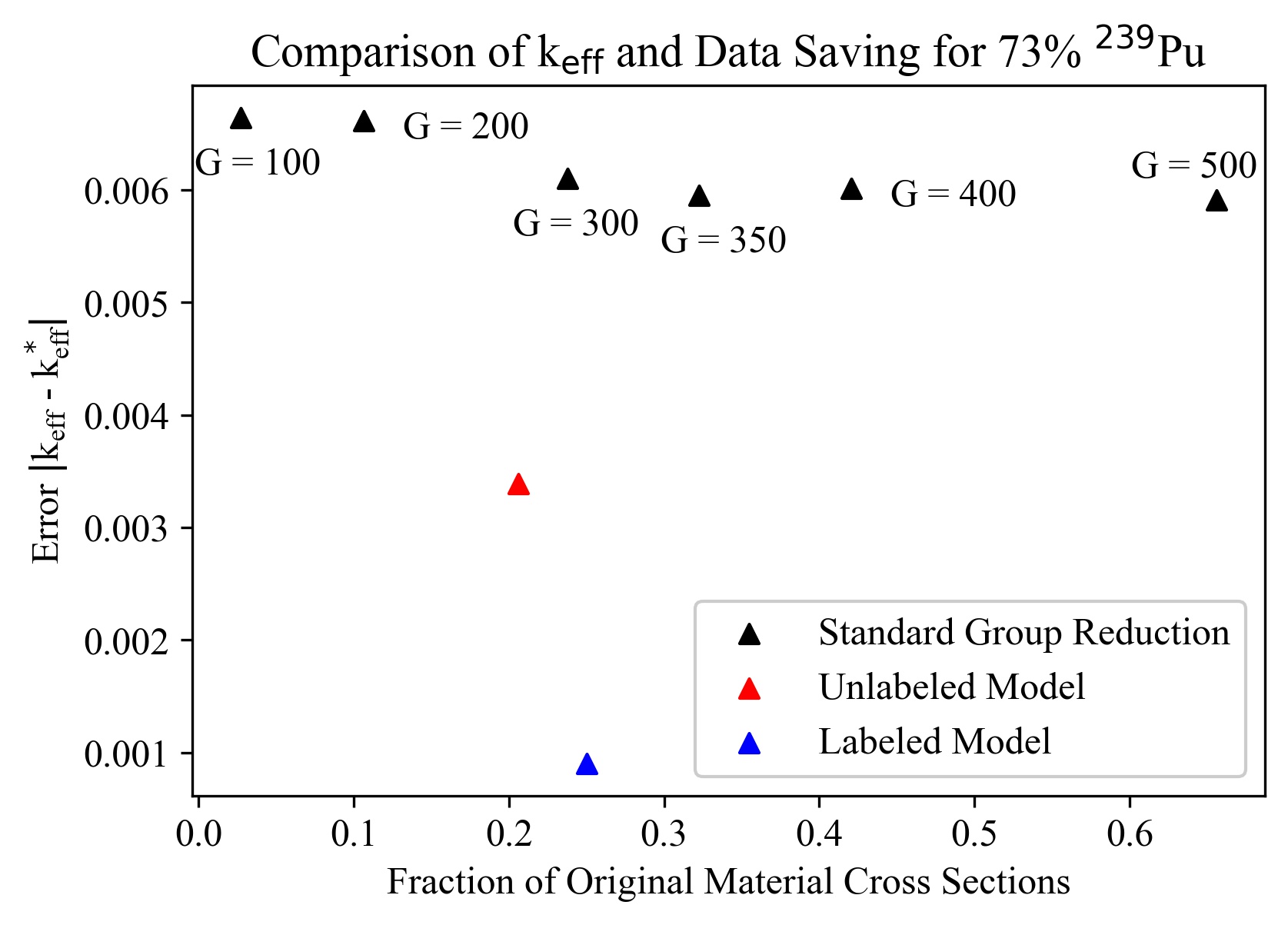}
    \caption{Comparison of the Data Saving Results for Standard Energy Group Reduction against the DJINN-autoencoder method. The data for the group reduction was comprised of the lower triangular scattering matrices and fission vectors of plutonium mixtures with $^{239}$Pu, 
    $^{240}$Pu, and HDPE. The reference $\keff$ value is the full model with $G = 618$ with a validation problem outside the training data range.}
    \label{fig:keff_vs_data}
\end{figure}

An alternative data saving technique would be to reduce the number of energy groups from $G = 618$ to a smaller number. 
Results for $G = 100$, $200$, $300$, $350$, $400$, and $500$ for the minimal storage case can be seen in Fig.~\ref{fig:keff_vs_data} against the ML models. 
The groups were coarsened in a mathematical manner so the new energy bins were made up of a similar number of energy bins.
Simply put, the $G = 100$ grid coalesced 6 and 7 bins of the $G = 618$ energy grid, with the 7 being evenly distributed through the new energy grid.
These data saving measures were compared against a problem with a plutonium mixture with 73\% $^{239}$Pu and the difference between the k-effective was plotted. 
Although $G = 100$, $200$, and $300$ used less data storage, they performed worse in terms of accuracy compared to the $G = 618$ reference value. 

\subsection{Perturbations to the Original Problem}
\begin{figure}[!ht]
    \centering
    \subfigure[]{\includegraphics[width=0.495\textwidth]{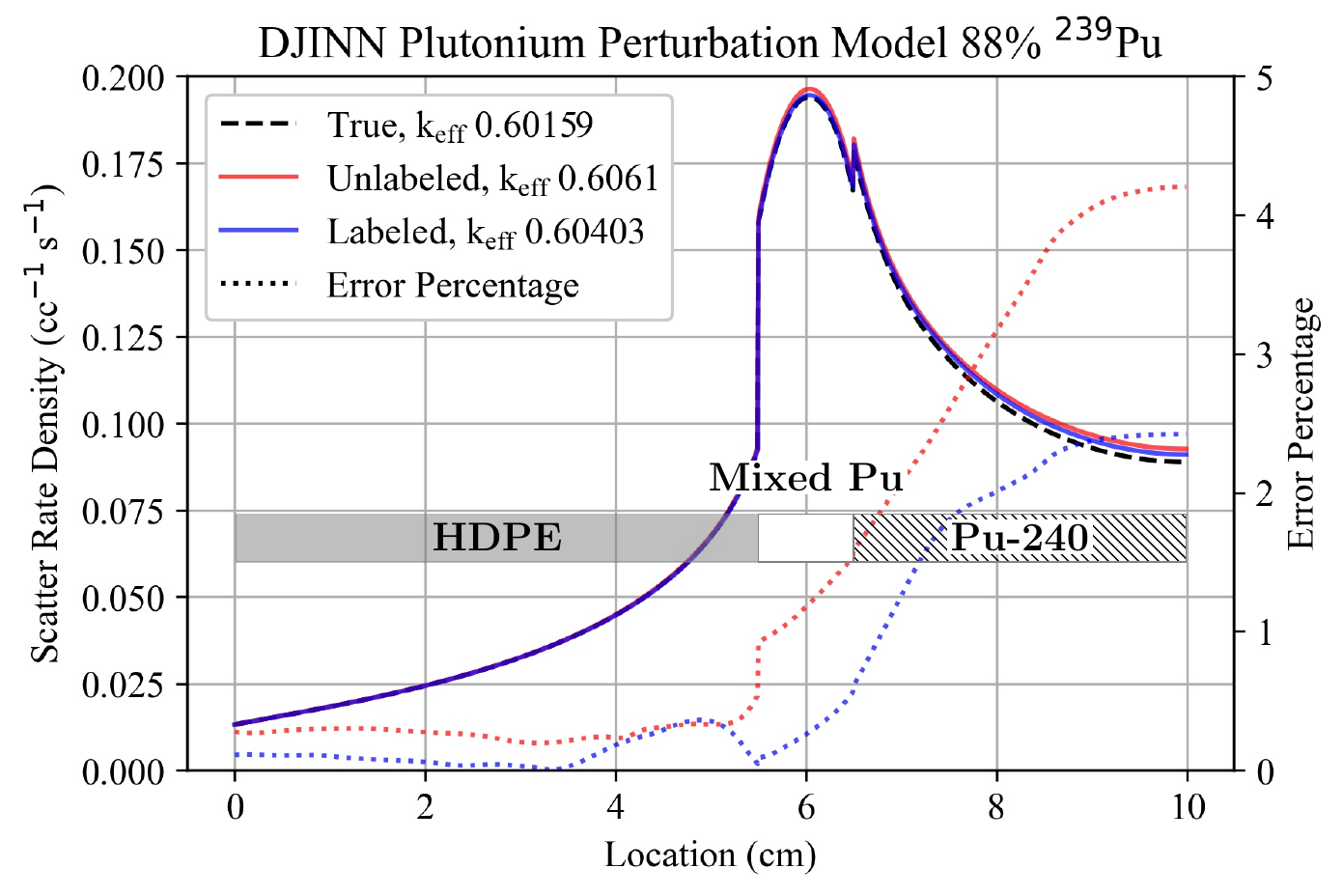}\label{fig:pluto_perturbation_results_12}}
    \subfigure[]{\includegraphics[width=0.495\textwidth]{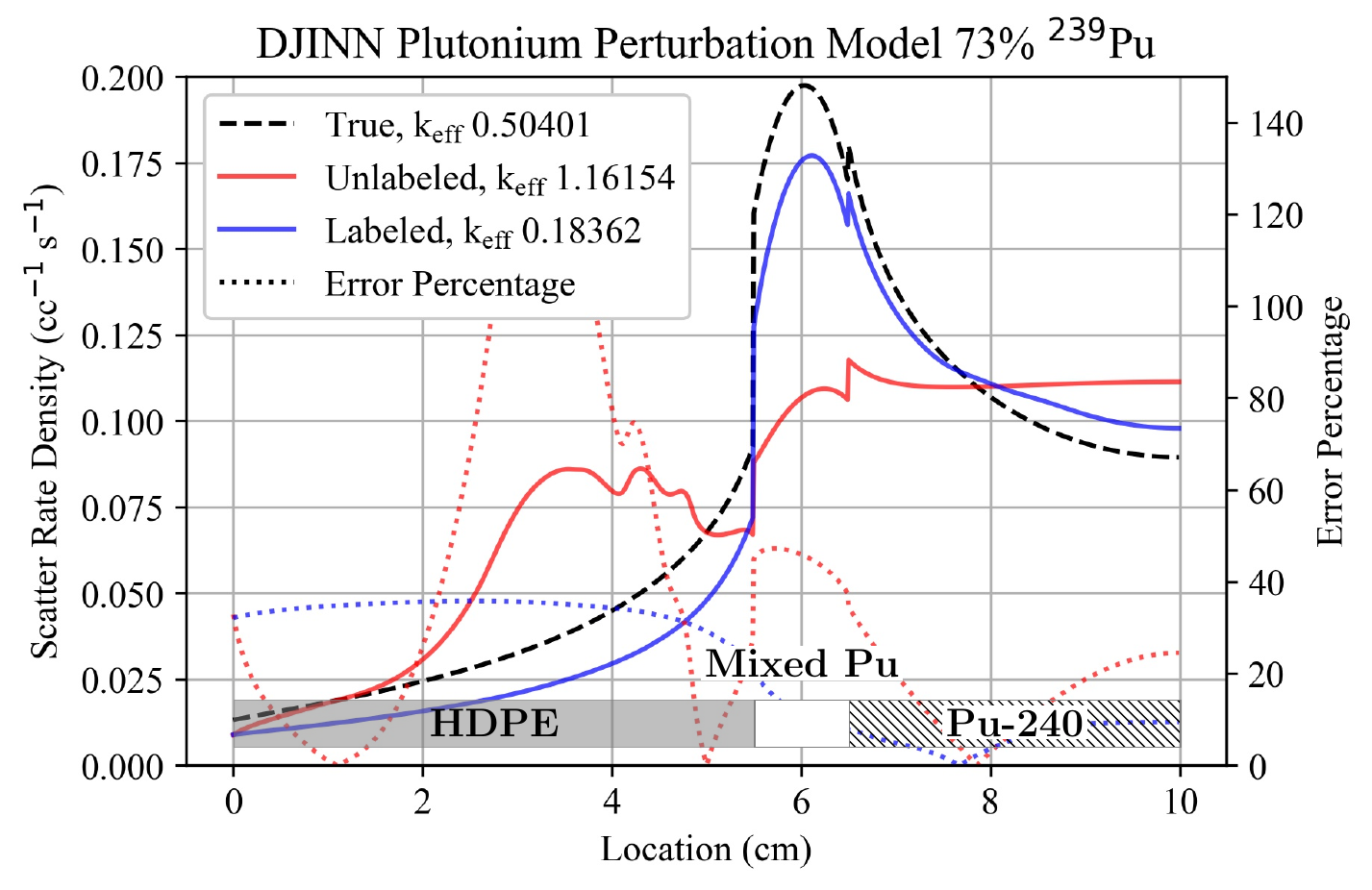}\label{fig:pluto_perturbation_results_27}}
    \caption{Scattering Rate Density results for the perturbed orientation \edit{(shown)} using the DJINN-autoencoder plutonium scattering and fission models and the DJINN-autoencoder HDPE model. \edit{The relative error associated with each model is also shown. It should be noted that the error in (a) and (b) use different scales.} (a) Shows a mixture of $^{239}$Pu inside the training data range where the ML models performs well. (b) Shows a mixture of $^{239}$Pu outside the training data in which there is inaccuracy with the ML models.}
    \label{fig:pluto_perturbation_results}
\end{figure}

While the ML models are able to predict the matrix-vector multiplication for different mixtures, they can also be used for different material widths of the original problem.
The perturbations to Fig.~\ref{fig:618_group_setup} are performed so the total width remains the same but the widths of two of the materials are altered by 0.5 centimeters.
All six of the perturbations are displayed in \ref{appendix:A}. 
Each of these perturbations were run with plutonium mixtures of 88\%, 85\%, and 73\% $^{239}$Pu, to show that it can be accurate both inside and outside the training data range.
The ML models accurately predicted the matrix-vector multiplication for all 88\% and 85\% $^{239}$Pu mixtures, as seen in Fig.~\ref{fig:pluto_perturbation_results_12}.
In this figure, the perturbation used decreases the width of the plutonium mixture region by 0.5 cm and increases the reflective material width by 0.5 cm\edit{, as shown in Fig.~\ref{fig:pluto_perturbation_results} and \ref{appendix:A}} ($^{240}$Pu Constant - 2).
\edit{The relative error in this figure shows that, although there is a different orientation of the problem, the DJINN-autoencoder models were  able to remain accurate.
It also shows that the error is increasing when comparing  the results in Fig.~\ref{fig:pluto_validation_results_12} to this figure.}

There were issues when the 73\% $^{239}$Pu mixtures were used in conjunction with a reduced width of the plutonium mixture material dropped to 1 centimeter. 
This is seen in Fig.~\ref{fig:pluto_perturbation_results_27}, which uses the same dimensions as Fig.~\ref{fig:pluto_perturbation_results_12}, but is unable to correctly predict the matrix-vector multiplication.
To verify the smaller plutonium mixture region was the cause of this inaccuracy, the plutonium mixture width in the two cases was set at 1 cm (Reflector Constant - 1 and $^{240}$Pu Constant - 2 in \ref{appendix:A}) were increased.
If the width of the plutonium mixture was 1.3 cm, the ML models were able to accurately predict the matrix-vector multiplication.
A width of 1.1 cm resulted in inaccuracies with the 73\% $^{239}$Pu mixture test problem.
\edit{With NNs, it becomes increasingly difficult to extrapolate the data outside the training data range~\cite{mccartney:2020}, as shown when varying both the material composition and the geometry of the problem.
This is similar to Fig.~\ref{fig:hdpe_multi_uh3_hdpe_models}, which shows the degradation with the DJINN models.
While the generality is shown with the DJINN-autoencoder models, there are limitations to this extrapolation.}

\subsection{Decrease in Wall Clock Time as Compared to Conventional Methods}
An additional benefit of replacing the scattering and fission cross section matrices with the DJINN-autoencoder models is the time required to run these problems, the wall clock time, was reduced.
\edit{This demonstrates that the replacement of the scattering and fission cross section matrices with ML models will alleviate the storage burden and results in faster convergence times.}
The initial method used in the S$_N$ transport code to store the different cross sections was to store the scattering and fission matrices at each spatial cell, and will be referred to as the spatial cross sections.
This is not an effective method when running with a significant number of spatial cells so it was expected that executing a problem with the DJINN-autoencoder was going to use less wall clock.

\begin{figure}[!ht]
    \centering
    \subfigure[]{\includegraphics[height=5.5cm]{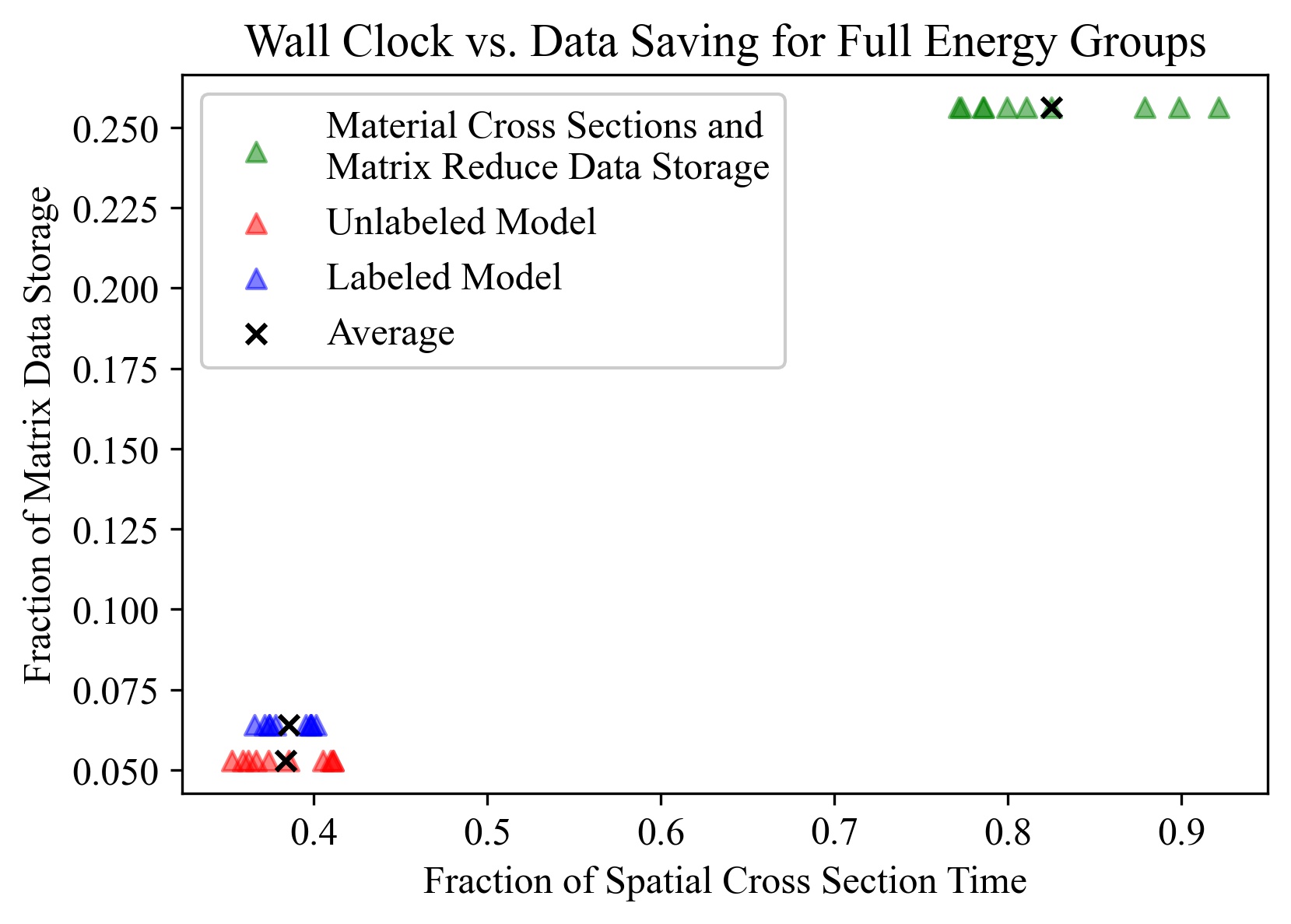}\label{fig:wall_clock_618}}
    \subfigure[]{\includegraphics[height=5.5cm]{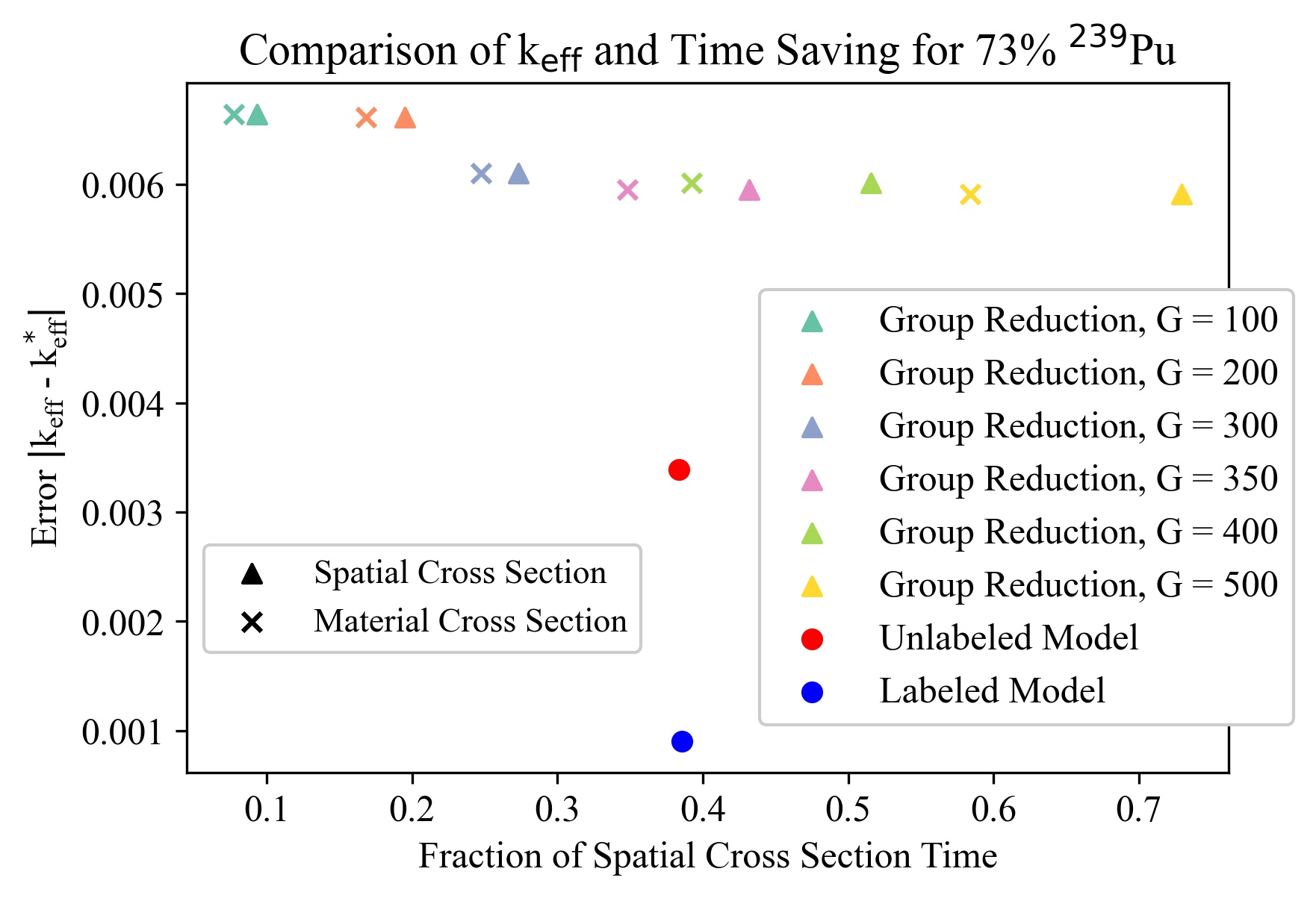}\label{fig:wall_clock_reduce}}
    \caption{Comparison of the wall clock times with the DJINN-autoencoder models and conventional methods. (a) Compares the labeled and unlabeled DJINN-autoencoder models for both wall clock time and the data storage requirements. These are compared to the material cross section wall clock time and the matrix reduce data storage. (b) Compares the DJINN-autoencoder labeled and unlabeled model times to those for a reduced number of groups. While the $G < 350$ autoencoder have faster wall clock times, the ML models get closer results to the reference $\keff$ value.}
    \label{fig:wall_clock}
\end{figure}

There were similar results when the ML method was compared to one where only the minimal number of cross sections matrices were stored in the transport code, one for each material.
This will be referred to as the material cross section.
Results comparing the material cross section against the ML models can be seen in Fig.~\ref{fig:wall_clock_618}.
The wall clock times and storage requirements were compared to the spatial cross sections and the full matrices.
Both the labeled and unlabeled DJINN-autoencoder models had quicker wall clock times and used less data space than the conventional methods.

To represent the wall clock speedup, the number of energy groups were reduced from $G = 618$ to sizes of $G = 100$, $200$, $300$, $350$, $400$, and $500$.
These coarser energy grid problems were then used to simulate a problem with a plutonium mixture with 73\% $^{239}$Pu. 
The results comparing these times against the spatial cross section wall clock time are shown in Fig.~\ref{fig:wall_clock_reduce}, which point out the speedup for both instances, the spatial and material cross sections.
Each of the data points were the mean values of ten simulations and plotted against the error in the k-effective value.
Although there is more of a speedup with the coarser grids as compared to the ML methods, the DJINN-autoencoder models are closer to the reference k-effective value in all instances.
\section{CONCLUSIONS AND FUTURE WORK}
The application of machine learning to replace the matrix-vector multiplication in the solution of neutronics problems is shown to be beneficial.
The data required to store the DJINN-autoencoder models used only 6.410\% of the data required to store the scattering and fission matrices of the different materials. 
These ML models still only used 25.015\% of the data when the upscattering was removed from the data and the fission matrix was reduced to its $\chi$ and $\nu \Sigmaf$ vectors.
In addition, the ML models provided generality for the materials and the widths of each of the materials. 
An unintended consequence was the speedup in the wall clock time, in which the ML models were able to iterate faster and use less data storage than the conventional methods. 
These speedups were similar to those of standard group reduction, however, the DJINN-autoencoder models were able to remain more accurate when it came to the k-effective value.

It should be pointed out that the original work, in which only DJINN was used, had materials with a lower number of energy groups, $G = 87$.
This fact is important as these ML models required 59.318\% of the original data storage as compared to the full matrices. 
This shows the advantages to using the DJINN-autoencoder models, in place of the scattering and fission matrices, are better when there are a large number of energy groups.
If decreasing the number of energy groups, there approaches a specific point in which conventional methods would be preferred to the machine learning methods. 

The next logical step in this process would be to obtain the highest performance from the autoencoders while minimizing their data space.
Tuning the hyperparameters, determining the optimal number of layers and nodes, and pruning the models should assist in addressing some of the areas that the ML models can improve, such as Fig.~\ref{fig:pluto_perturbation_results_27}.
While tuning was used for some aspects of the autoencoder training, not all hyperparameters have been examined closely for achieving the optimal latent space to allow for DJINN training. 

\edit{While this application only used one dimensional slab geometry, it would be better served with more realistic models.
The application of DJINN and DJINN-autoencoder models on multi-dimensional problems should be pursued.
This will verify that the implementation of ML models is beneficial for more realistic systems.}

Furthering the generality of the ML models can be seen with the addition of a temperature dependence and anisotropic scattering.
While it was briefly mentioned that both of these variables will impact the scattering and fission matrices, no work has been completed to date as the problems have been isotropic and used a single temperature.
It would also be beneficial if there was a method to combine multiple reflective materials into a single ML model.
Preliminary tests have been done but the ML models show a degradation in performance.

Potential ground for future work includes using the idea of machine learning to reduce the memory requirements for continuous energy Monte Carlo simulations of neutron transport. The description of continuous energy physics parameters could be ripe for similar data reduction. Additionally, the development ``on-the-fly'' methods for constructing the neural networks should also be explored. For instance, in a design study random samples from the design space could be used to generate training data and then a trained model could be applied to other regions of space.

\section*{Acknowledgement}
This work was supported by Lawrence Livermore National Laboratory under subcontract B639472, ``Improved Nuclear Data for Transport Simulations using Machine Learning'' and by the Center for Exascale Monte-Carlo Neutron Transport (CEMeNT) a PSAAP-III project funded by the Department of Energy, grant number DE-NA003967.

\setlength{\baselineskip}{12pt}
\bibliographystyle{elsarticle-num}
\biboptions{numbers,sort&compress}
\bibliography{resources}

\newpage
\appendix
\section{Different Perturbations of the 618 Group Problem} \label{appendix:A}

\begin{figure}[!h]
    \begin{center}
	\includegraphics[page=1]{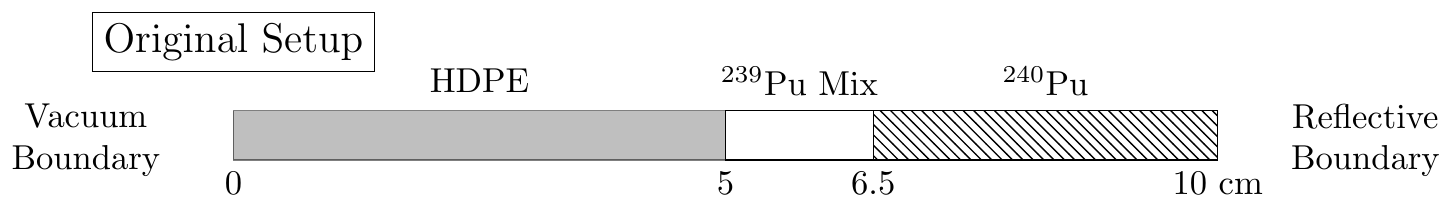}
	\includegraphics[page=2]{figures/g618_tikz/plutonium_perturbs.pdf}
	\includegraphics[page=3]{figures/g618_tikz/plutonium_perturbs.pdf}
	\includegraphics[page=4]{figures/g618_tikz/plutonium_perturbs.pdf}
	\includegraphics[page=5]{figures/g618_tikz/plutonium_perturbs.pdf}
	\includegraphics[page=6]{figures/g618_tikz/plutonium_perturbs.pdf}
	\includegraphics[page=7]{figures/g618_tikz/plutonium_perturbs.pdf}
        \caption{Different perturbations of the original Plutonium 618 Group Problem}
        \label{fig:perturbations}        
    \end{center}
\end{figure}

\end{document}